\documentclass[aps,amsmath,twocolumn,amssymb,floatfix,showpacs,superscriptaddress,nofootinbib,longbibliography]{revtex4-1}
\usepackage{amsmath}
\usepackage{mathtools}
\usepackage{braket}
\usepackage{graphicx}
\usepackage{amssymb}
\usepackage[dvipsnames]{xcolor}
\usepackage{multirow}
\usepackage{xfrac}
\usepackage{textcomp}
\usepackage{float}
\usepackage{enumerate}
\usepackage{subfigure}
\usepackage{booktabs}
\usepackage[colorlinks=true,linktoc=page,linkcolor=blue,citecolor=blue,urlcolor=blue]{hyperref}
\usepackage{verbatim}

\mathchardef\mhyphen="2D 

\usepackage[normalem]{ulem}
\definecolor{lime}{HTML}{A6CE39}
\usepackage{sidecap,tikz}
\DeclareRobustCommand{\orcidicon}{\hspace{-1.0mm}
	\begin{tikzpicture}
	\draw[lime, fill=lime] (0.0,0.0) 
	circle [radius=0.15] 
	node[white] {{\fontfamily{qag}\selectfont \tiny \,ID}};
	\draw[white, fill=white] (-0.0525,0.095) 
	circle [radius=0.007];
	\end{tikzpicture}
	\hspace{-3.0mm}
}
\foreach \x in {A, ..., Z}{\expandafter\xdef\csname orcid\x\endcsname{\noexpand\href{https://orcid.org/\csname orcidauthor\x\endcsname}
		{\noexpand\orcidicon}}
}

\begin{document}

\renewcommand{\thefootnote}{\fnsymbol{footnote}}


\title{Deciphering competing interactions of Kitaev-Heisenberg-\texorpdfstring{$\Gamma$}~~system in clusters:\\part I - static properties}	

\author{Sheikh Moonsun Pervez\orcidA{}}
\email{moonsun@iopb.res.in}
\affiliation{Institute of Physics, Sachivalaya Marg, Bhubaneswar-751005, India}
\affiliation{Homi Bhabha National Institute, Training School Complex, Anushakti Nagar, Mumbai 400094, India}

\author{Saptarshi Mandal\orcidB{}}
\email{saptarshi@iopb.res.in}
\affiliation{Institute of Physics, Sachivalaya Marg, Bhubaneswar-751005, India}
\affiliation{Homi Bhabha National Institute, Training School Complex, Anushakti Nagar, Mumbai 400094, India}
\begin{abstract}
Recently, the Kitaev-Heisenberg-$\Gamma$ system has been used to explore various aspects of Kitaev spin liquid physics. Here, we consider a few small clusters of up to twelve sites and study them in detail to unravel many interesting findings due to the competition between all possible signs and various magnitudes of these interactions under the influence of an external magnetic field. When Heisenberg interaction is taken anti-ferromagnetic, one obtains plateaus in correlation functions where, surprisingly, the exact groundstate reduces to the eigenstate of Heisenberg interaction as well. On the other hand, for ferromagnetic Heisenberg interaction, its competition with Kitaev interaction results in non-monotonicity in the correlation functions. We discuss, in detail, the competing effects on low energy spectrum, flux operator, magnetization, susceptibility, and specific heat. Finally, we discuss how our findings could be helpful to explain some of the recent experimental and theoretical findings in materials with Kitaev interactions.
\end{abstract}
\date{\today}
\maketitle
\section{Introduction}\label{intro}
From its inception, the Kitaev model\cite{kitaev-2006} has attracted unprecedented interest from condensed matter community for its simplicity and yet exact realization of many unique properties such as fractionalization, topological order, anyonic excitations etc.  Though initially proposed as a mere toy model with certain aspects of fault-tolerant quantum computation in perspective, it has  been investigated to examine many other aspects of many-body condensed matter physics. This includes, for example, study of spin-spin correlation~\cite{smandal-2007,tikhonov-2011,lunkin-2019}, effect of disorder and localization~\cite{kao-2021,nasu-2020}, Kondo effects and its competition with Kitaev physics~\cite{tathagata-2020}, confinement-deconfinement transition and quantum phase transitions in perturbed Kitaev model~\cite{saptarshi-2011,animesh-2020,animesh-2021, knolle-2018}, entanglement \cite{yaoentanglement,mandalentanglement} etc. Variety of lattices in two and three dimensions with Kitaev-like interaction \cite{kivelson-2007,vala-csl-2010,naveen-2008,eschmann-2020}, showed exciting features, such as the existence of a chiral spin liquid, gapless Fermi contour, etc. Also, the order-disorder mechanism has been studied in the Kitaev model for classical spins~\cite{baskaran-2008,samarth-2010}. The emergence of orbital magnetization in Kitaev magnets~\cite{saikat-lin}, and the possibility of tuning the exchange parameters by using light to stabilize the spin liquid phase~\cite{Kumar2022}, observe inverse Faraday effect~\cite{saikat-inverseFARADAY} have been explored. Apart from these theoretically motivated works, it is now established that there are several materials proposed to possess Kitaev-like interaction along with other kinds of interactions ~\cite{chalaupka-2010,simon-2022,takagi-2019,motome-2020,hermanns-2018,abanerjee-2016}. The presence of non-Kitaev interactions makes it challenging to realize the Kitaev physics exactly and efforts are underway to understand the full scope of such interactions \cite{janssen-2017,chern-2021,czajka-2021,niravkumar-2019,wulferding-2020,berke-prb-2020,haoxiang-nature-2012,balz-prb-2021,plessis-2020,sananda-2019,sananda-2021}.
\\\\\indent
One fascinating aspect of condensed matter systems is that physical interactions at the microscopic range govern the physical properties at macroscopic length \cite{anderson,fifty-years,savary-review}. In the Kitaev model, the bond-dependent short-range correlation function and spin-fractionalization are realized precisely even for the smallest size if appropriately defined. This realization renders some aspects of Kitaev magnetism for the Kitaev model in small clusters \cite{tathagata-2020,kataoka-jpsj-2020}. In this study, we consider the Kitaev-Heisenberg-$\Gamma$ ($K$-$J$-$\Gamma$) model \cite{Rau-2014} to investigate how various local exchange couplings compete with each other and whether there is a pattern in it leading to an in-depth understanding of this system in thermodynamic limit~\cite{wang-2020}. To our knowledge, such a detailed comparative exact study is absent within a single work. To this end, we consider an exact analysis of $K$-$J$-$\Gamma$ systems in small clusters and investigate some very essential quantities such as eigenspectrum, nearest-neighbour spin-spin correlation function, flux operators, magnetization, susceptibility and specific heat for all possible sign combinations of the interactions, under the influence of external magnetic field ($h_z$) taken in $\hat{z}$ direction. We hope our study will serve as a valuable reference for the $K$-$J$-$\Gamma$ system.
\\\\\indent
Our plan of presentation is as follows. In Sec.\ref{spectrum}, we show how low energy eigenvalues vary as a function of Heisenberg strength ($J$) and $h_z$. Our analysis shows that eigenenergies of the $K$-$J$-$\Gamma$ system can be grouped into two branches. In the first branch, energies vary linearly for all ranges of $J$. For the second branch, energies vary linearly only for large magnitude of $J$, but for small values, they show a cusp or dip near $J=0$ whose exact position depends on the value of $\Gamma$ and cluster size. Interestingly, the linear branches change their energies from positive to negative when the sign of $J$ is reversed, but for the second type, it remains negative or positive as we change the sign of $J$. The incidental crossing between the two types of energies can be controlled by tuning $\Gamma$. Afterwards, we focus on the nearest-neighbour correlation function in Sec.\ref{correlation} and study how it depends on cluster size and different system parameters as before. We have considered both $\sigma^z_i \sigma^z_j$ and $\sigma^y_i \sigma^y_j$, on a `$z$-type' ($z$-type - in pure Kitaev limit) bond. The correlation function shows interesting variations depending on the system parameters' relative sign and magnitude. Interestingly, we distinguish clearly between the ferromagnetic (FM) and anti-ferromagnetic (AFM) $J$ on the nearest neighbour correlation function. For AFM $J$, the correlation function shows plateau-like behaviour where spin-spin correlation does not change as we change $J$ and corresponds to a fixed point where the exact eigenstate also becomes the eigenstate of Heisenberg interaction. Meanwhile, for FM $J$, the correlation function shows non-monotonicity, suggesting a competing mechanism induced by $K$ and FM $J$. The fate of the conserved plaquette operator is discussed briefly in Sec.\ref{bpoperator}.
\\\\\indent
Finally, we present finite temperature magnetization, susceptibility, and specific heat in Sec.\ref{magnetization}, \ref{susceptibility}, and \ref{specificheat}, respectively. We describe how these quantities depend on various magnetic exchange coupling, including external field. The primary finding on magnetization is that, for AFM $K$ and low magnetic field, there is an intermediate range of temperature where magnetization possesses a smooth dome-like behaviour, which is absent for FM $K$. However, depending on the $J$, this dome-like structure can be re-entrant for FM $K$. The external magnetic field is shown to control the zero temperature magnetization, which also depends differently for different combinations of $J,~K,~\Gamma$, and cluster size. Like magnetization, susceptibility, and specific heat have distinct features for FM and AFM $K$. Again, they can change differently depending on the sign and magnitude of the $J$ and the applied external magnetic field. We conclude our findings in Sec.\ref{discussion} with a comparison to recent experiments and other theoretical studies.
\section{Kitaev-cluster-Basics and spectrum analysis}\label{spectrum}
\begin{figure}[h]
	\centering
	\includegraphics[width=0.85\columnwidth,height=!]{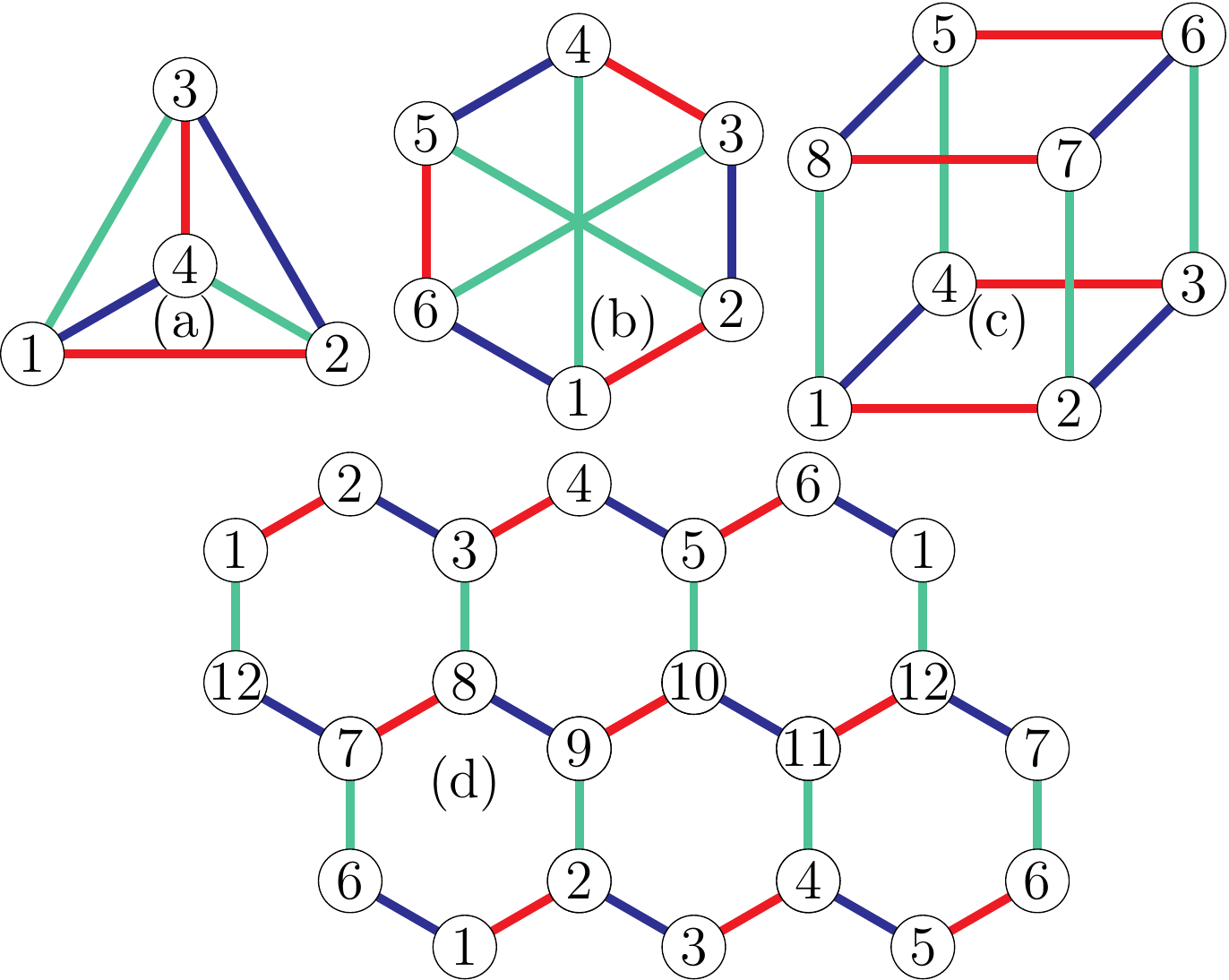}
	\caption{Depiction of (a) 4-site, (b) 6-site, (c) 8-site, and (d) 12-site Kitaev cluters. A bond's red, blue, and green colour indicates $x,~y,$ and $z$-type Ising interaction (in the pure-Kitaev limit) between the two sites holding that particular bond.}
	\label{clusterfig}
\end{figure}

\begin{figure*}
	\includegraphics[width=2.00\columnwidth,height=!]{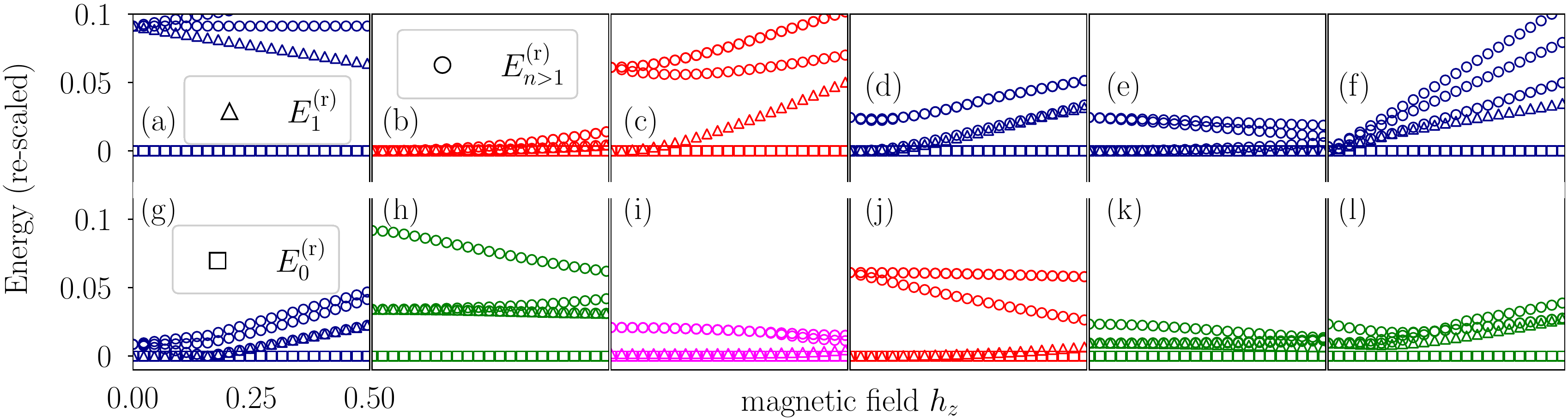}
	\caption{Few lowest eigenvalues are plotted with $h_z$ in absence of $\Gamma$. For better visualization, eigenvalues are shifted and re-scaled as $E_n^{({\rm r})}=\frac{(E_n-E_0)}{6N}$; $N$ is the cluster size. Red, blue, green, and magenta lines are for 4, 6, 8, and 12-site clusters. Different points' shapes indicate different energy levels, as shown in the legends. We have twenty-four plots for four types of clusters, with AFM and FM Kitaev interaction, for $J=-0.5,0.0,+0.5$. However, here, we plot only twelve out of them, and there is only one representative panel for similar-looking panels. The $(N,K,J)$ values are (a) (6,1,0.5), (8,1,0.5), (12,1,0.5), (b) (4,-1,0.5), (6,-1,0.5), (8,-1,0.5), (12,-1,0.5), (c) (4,-1,0), (4,1,-0.5), (4,-1,-0.5), (d) (6,-1,0), (12,-1,0), (e) (6,1,0), (12,1,0), (f) (6,-1,-0.5), (8,-1,-0.5), (12,-1,-0.5), (g) (6,1,-0.5), (h) (8,1,-0.5), (i) (12,1,-0.5), (j) (4,1,0), (4,1,0.5), (k) (8,1,0), (l) (8,-1,0). In the panel, we plotted the first member of each of these groups.}
	\label{eigenvalue-in-hz}
\end{figure*}

The defining Hamiltonian we investigate in our study is $K$-$J$-$\Gamma$ model~\cite{wang-2021}, which is given as,
\begin{eqnarray}
	H= \sum_{\langle i,j \rangle_{\gamma\neq\alpha
			\neq \beta}} \left( K S^{\gamma}_iS^{\gamma}_j +J~{\bf S}_i \cdot {\bf S}_j + \Gamma(S^{\alpha}_iS^{\beta}_j + S^{\beta}_iS^{\alpha}_j) \right),~~~~
\end{eqnarray}
where nearest neighbours $\langle i,j\rangle$ are connected by $\gamma$-type ($\gamma=x,\text{ or }y,\text{ or }z$) bond in pure Kitaev limit. In addition, we have considered an external Zeeman field along $z$-direction. We have taken 4, 6, 8, and 12-site clusters for our study, as depicted in FIG.\ref{clusterfig}.
We have used the exact diagonalization technique to obtain the energy eigenvalues and eigenstates of these clusters for various parameter values. We have used these obtained state(s) to calculate various zero and finite temperature properties. We also note that the 6-site cluster considered here coincides with the hexagonal-plaquette taken previously~\cite{alternative_6_site} (containing next to next nearest neighbour interaction) after exchanging a pair of sites or by local gauge transformations. Thus, it relates to a physically motivated extended Kitaev model.
\\\\\indent
Experimentally, it has been found that the Zeeman field brings in a non-trivial effect as far as the spin-liquid phase is concerned \cite{berke-prb-2020,nasu-motome-2015,koga-nasu-2019,li-2021-ncom}.
\begin{figure*}\centering
	\includegraphics[width=2.00\columnwidth,height=!]{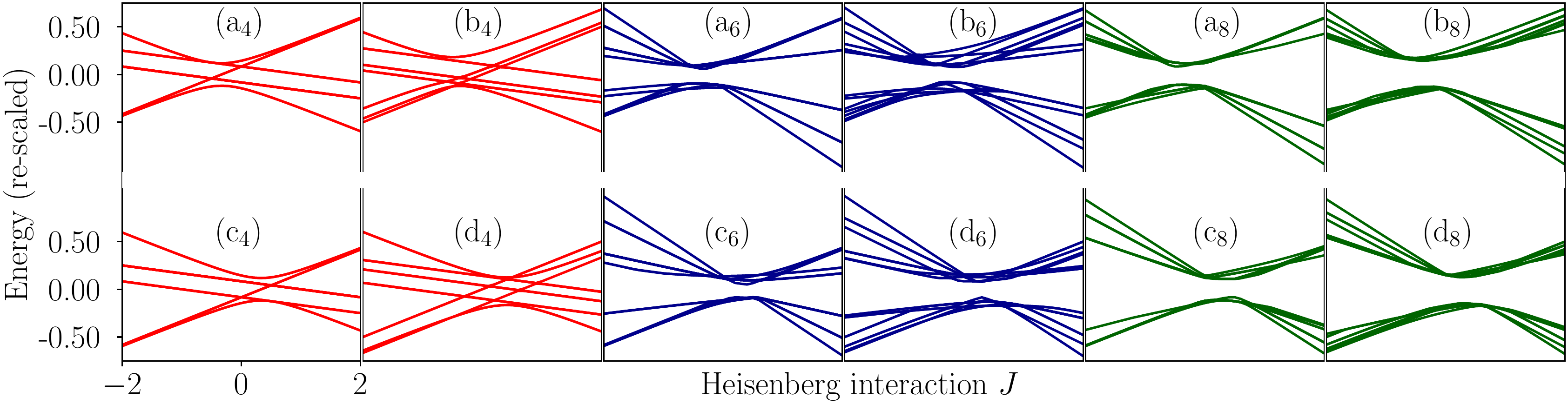}
	\caption{Eigenvalues are plotted for various strengths of $J$. For better visualization, eigenvalues are divided by $6 N$ ($N:$ cluster size). Lines in red, blue, and green are for 4-site, 6-site, and 8-site clusters. Parameter values ($K,\Gamma$) in the plots are taken as (a$_N$) 1.0, 0.0, (b$_N$) 1.0, 0.5, (c$_N$) -1.0, 0.0, (d$_N$) -1.0, 0.5. For $N>4$, only the lowest and highest ten energy levels are shown.}
	\label{eigenspectrum}
\end{figure*}
In FIG.\ref{eigenspectrum}, we have plotted the eigenvalue spectrum for various clusters for different choices of $J, K, \Gamma$. We first discuss the simple case of $\Gamma=0$. We notice that eigenvalues can be grouped into two categories: the one whose eigenvalues vary linearly with $J$ and the one whose does not vary linearly with $J$. This behaviour is true for all clusters. Interestingly, a finite $\Gamma$ affects the eigenspectra in two ways. Firstly, it breaks the degeneracy of eigenvalues (of the first kind) found for $\Gamma=0$. Secondly, the bandwidth increases for non-zero $\Gamma$ and higher for positive $\Gamma$. Also, $\Gamma$ determines the gap (separation between the ground and the first excited state's energy) at $J=0$.
\\\\\indent 
As we decrease the value of $J$ from a substantial positive value to a large negative value, the quantum ground state is continually adjusted from a large singlet component to a large FM component for $\Gamma=0$. However, for non-zero $\Gamma$, as we go from large AFM $J$ to large FM $J$, there is a crossover between the states $E_{0}$ to $E^n$ at some finite $J$. Here $E_0$ is the lowest energy state at $J \rightarrow - \infty$, and $E^{n}$ is some positive high energy states at $J \rightarrow + \infty$. These two states continually come closer to each other, and for $\Gamma=0$, they asymptotically merge, but for finite $\Gamma$, there is a crossover between these two states at some critical value of $J$.
\\\\\indent
For the sake of simplicity, we first start analyzing the effect of the external magnetic field for $J=0$ and $J=-0.5$
as shown in FIG.\ref{eigenvalue-in-hz}.
We observe these two choices of
$J$, when $K$ is FM, the eigenvalues mainly increase with the magnitude of $h_z$. However, for AFM $K$, the eigenvalues decrease or increase depending on the cluster size and strength of the external magnetic field. Notably, in the 12-site cluster, for $(K, J)=(-1,-0.5)$ (panel (f)), we observe that the gap between the ground state and the first few excited states is gradually increasing with $h_z$. But, for AFM $K$,
i.e., for $(K, J)=(1,-0.5)$ (panel (i)), the ground state and the first excited state remain nearly degenerate. Also, in the former case, the other excited states move away from the ground state, but in the latter case, the low-energy excited states move closer toward the ground state energy. This highlights that the relative signs and magnitudes of $K$ and $J$ compete very differently. A similar analysis can also be done for AFM $J$. It is interesting to compare our result to that found in an earlier study~\cite{zhu-2018} where the effect of an external magnetic field in the direction of $\left(1,1,1\right)$ has been considered. FM and AFM $K$ have been shown to produce qualitatively different results. With FM $K$, the first and other excited states show a monotonous increase of the gap from ground state energy, which we also observe as evident from panel (c,d,l) of FIG.\ref{eigenvalue-in-hz}. However, for AFM $K$, a qualitatively similar evolution of eigenenergies has been observed, where higher excited states come closer to the ground state energy. At some critical external field, it nearly degenerates with the ground state \cite{zhu-2018}. From FIG.\ref{eigenvalue-in-hz}, we observe a qualitatively different evolution of low-lying eigenstates with the external magnetic field than that found for FM Kitaev interaction. However, the difference between earlier result~\cite{zhu-2018} and in our case is attributed to finite size\cite{holzmann-2016} and direction of the magnetic field, which has been taken in $\left(0,0,1\right)$ direction.
\section{Correlation function}\label{correlation}
In pure Kitaev limit, the two-spin correlation functions are bond dependent (but independent of system sizes) and exist for nearest-neighbour only, a consequence of $Z_2$ conserved operators. Meanwhile, in a pure $J$ limit, spin-spin correlation functions for nearest neighbour spins are isotropic because of $SU(2)$ symmetry. The characteristic bond-dependent correlation function in the Kitaev limit does not survive when any other interaction, such as $J$, $\Gamma$, or $h_z$, is applied. Here we look for the manifestations of other non-Kitaev interactions in a $K$-$J$-$\Gamma$ model\cite{khwang-2022,feng-2022,han-li-2023,yilmaz-2022,shuyi-2023,smit-2020,kimchi-2014,yamaji-2016,young-baek-kim-2015,wang-2021} in clusters of different sizes.
\subsection{$z$-$z$ correlation function}
\begin{figure*}\centering
	\includegraphics*[width=1.8\columnwidth,height=!]{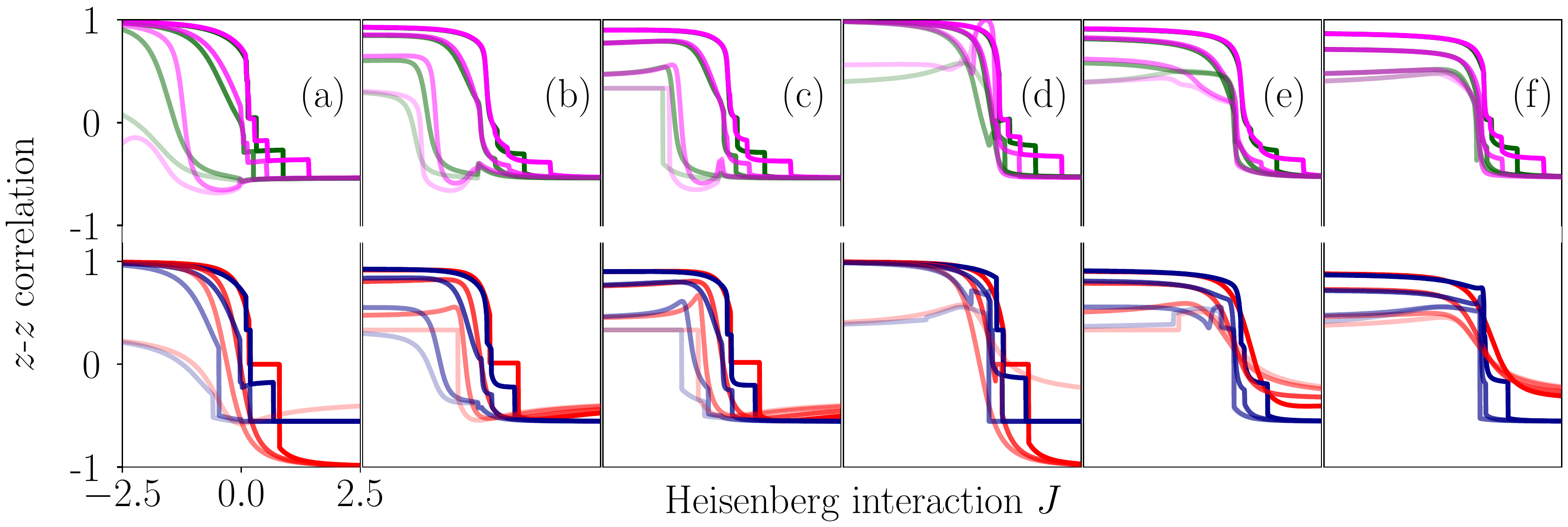}
	\caption{Nearest neighbor $z$-$z$ correlation on a $z$-type bond vs. $J$, for different $h_z$. $K$ is AFM in (a-c) and FM in (d-f). Value of $\Gamma$ is (a) 0.0, (b) 0.5, (c) -0.5, (d) 0.0, (e) 0.5, (f) -0.5. Red, blue, green, and magenta lines are for 4, 6, 8, and 12-site clusters. Lighter to darker shades indicate higher $h_z$ values; four different $h_z$ values have been used to plot: 0.0, 0.2, 1.0, and 2.0.}
	\label{zzcorrelation}
\end{figure*}

Without loss of generality, we consider the bonds $\left(1,3\right)$, $\left(1,4\right)$, $\left(1,8\right)$ $\left(1,12\right)$ for the four, six, eight, and twelve-site cluster. They are all `$z$-type' bonds (in the pure-Kitaev limit), as shown in FIG.\ref{clusterfig}. The $z$-$z$ correlations (CR$_{z-z}$) are shown in FIG.\ref{zzcorrelation}. To begin with, we first discuss the scenario $\Gamma=0$.
\\\\\indent

{\it Correlation function for $\Gamma=0$:} Let us first concentrate on the case for $\Gamma=0$ (panel (a) and (d) of FIG.\ref{zzcorrelation}), and $h_z=0$. There is a point of inflexion in ${\rm CR}_{z-z}$ at $J=0$. However, for the 4-site cluster, ${\rm CR}_{z-z}$ changes drastically with the external magnetic field and becomes independent of whether $K$ is FM or AFM (between them, the difference is merely a shift along the $J$-axis). In both cases, ${\rm CR}_{z-z}$ saturates to $\pm 1$ for large $\mp J$. As we increase $h_z$, after a critical $h_{z}^{(c)}$, ${\rm CR}_{z-z}$ becomes constant for a certain range of AFM $J$. The ground state does not change much, such that ${\rm CR}_{z-z}$ remains constant for a range of $J$. The above nature for ${\rm CR}_{z-z}$ changes as we increase the cluster size with two notable differences. Firstly, the zero and non-zero field ${\rm CR}_{z-z}$ for AFM $J$ saturates to the same value for $N\geq6$. Secondly, the number of plateau regions increases with cluster size. For FM $J$, ${\rm CR}_{z-z}$ monotonically increases and reaches a saturation whose upper bound is 1 for a very large $h_z$. The correlation plateau is universally observed for all clusters in a certain range(s) of AFM $J$, and the width of these plateaus increases with increasing $h_z$.
\\\\\indent
Now, we explain the origin of plateau-like regions for AFM $J$. Our exact numerics show that for some values of $K, J, \Gamma, h_z$, the ground state becomes a simultaneous eigenstate of Heisenberg interaction and the Zeeman term. This is intriguing because the Heisenberg interaction and external magnetic field do not commute with the Hamiltonian. For the four-site cluster, the ground state for the plateau region is given by,
\begin{eqnarray}\label{GS_in_plateau}
	| \mathcal{G} \rangle	&=&\frac{1}{2}\ket{\uparrow_{1}\uparrow_{2}} \left( \ket{\uparrow_{3}\downarrow_{4}}-\ket{\downarrow_{3}\uparrow_{4}}\right) \nonumber \\ 
	&& + \frac{1}{2}\left( \ket{\uparrow_{1}\downarrow_{2}}-\ket{\downarrow_{1}\uparrow_{2}} \right) \ket{\uparrow_{3}\uparrow_{4}}.
\end{eqnarray}
This wave function belongs to odd-spin sectors where the basis states involved are degenerate eigenstates of the Zeeman term. The Zeeman term generally introduces a diagonal term determined by the difference $N_{\uparrow}- N_{\downarrow}$ where $N_{\sigma}$ denotes that the total number of $\sigma$-polarizations in a given basis state. However, it may happen that for a set of basis states $N_{\uparrow}- N_{\downarrow}$ is identical as happens for \ref{GS_in_plateau}. Here, it appears that the ground state accidentally has a larger symmetry than the Hamiltonian itself.
\\\\\indent
As far as the non-monotonicity is concerned for the AFM Kitaev interaction and FM Heisenberg interaction, we note that the bond Hamiltonian in the absence of $\Gamma$ is $H^z_{ij}= K \sigma^z_i\sigma^z_j - |J| (\sigma^x_i\sigma^x_j + \sigma^y_i\sigma^y_j + \sigma^z_i\sigma^z_j)$. For convenience, we can always choose the z-axis along the direction of the first spin, which renders $H^z_{ij}= (K-|J|) \cos \theta $ where $\cos \theta=\sigma^z_i\sigma^z_j $. In the absence of $J$, the positive $K$ needs a $\theta_0 > \pi/2$. For finite and small $|J|$, the $\theta$ can maximally reach to $\pi$ to make $(K-|J|) \cos \theta$ minima. For large $|J|$, we find that $\theta \rightarrow 0$, and this explains the non-monotonicity qualitatively.
\begin{figure*}
	\includegraphics*[width=1.800\columnwidth,height=!]{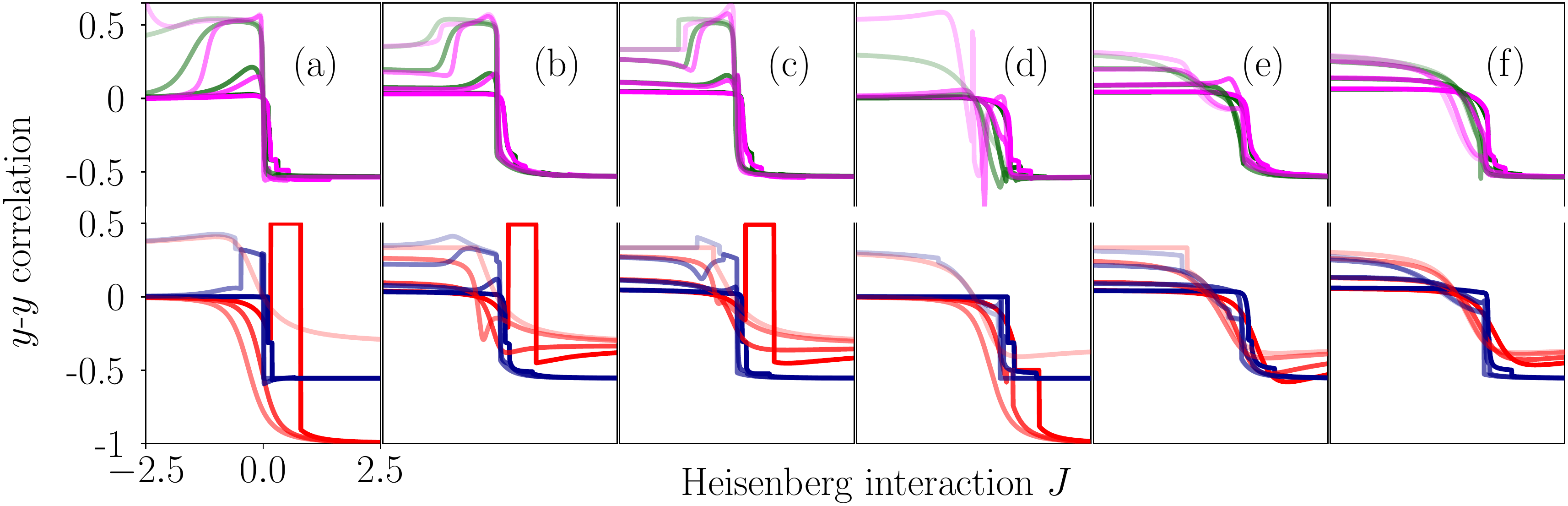}
	\caption{Nearest neighbor $y$-$y$ correlation on a $z$-type bond vs. $J$, for different $h_z$. $K$ is AFM in (a-c) and FM in (d-f). Value of $\Gamma$ is (a) 0.0, (b) 0.5, (c) -0.5, (d) 0.0, (e) 0.5, (f) -0.5. Red, blue, green, and magenta lines are for 4, 6, 8, and 12-site clusters. Lighter to darker shades indicate higher $h_z$ values; four different $h_z$ values have been used to plot: 0.0, 0.2, 1.0, and 2.0.}
	\label{yycorrelation}
\end{figure*}
\\\\\indent
Firstly, for finite $\Gamma$, the asymptotic value of ${\rm CR}_{z-z}$ for FM $J$ depends on the external magnetic field (in contrast to $\Gamma=0$ case) and increases with its strength. Secondly, for $K=1$, $\Gamma$ decreases ${\rm CR}_{z-z}$ for a small range of FM $J$, which is more prominent for the 12-site cluster.
\\\\\indent
We also note that for large AFM $J$ and finite $\Gamma$, for the four-site cluster, the asymptotic value of ${\rm CR}_{z-z}$ saturates to $-\frac{1}{3}$ (independent of $h_z$) and equals to $y-y$ and $x-x$ correlations as ground state reduces to singlet state represented by $\ket{\rm GS}= \left( \frac{1}{2}\left(\ket{\uparrow_1\downarrow_3}- \ket{\downarrow_1\uparrow_3}\right) \times \left( \ket{\uparrow_2\downarrow_4}-\ket{\downarrow_2\uparrow_4}\right)\right)$. For cluster size $N\geq 6$, ${\rm CR}_{z-z}$ saturates to $- \frac{5}{9}$ irrespective of values of $h_z$ and $\Gamma$.
We note that for $\Gamma=-0.5$, CR$_{z-z}$ is qualitatively similar to that of $\Gamma=0.5$, though there are some critical $J_c$
for which CR$_{z-z}$ shows maximum at some $J$ for larger clusters.\\
\indent
\subsection{$y$-$y$ correlation function} In the Kitaev limit, $y$-$y$ correlation function (CR$_{y-y}$) on a $z$-type bond has zero value. However, finite $J$ and $\Gamma$ interaction brings some non-zero values to CR$_{y-y}$ (FIG.\ref{yycorrelation}). When $\Gamma=0,~h_z=0$, CR$_{y-y}$ saturates to some constant positive (negative) value for substantial FM or AFM $J$.
There is some non-monotonicity due to the competition between Kitaev and $J$ for intermediate FM $J$ values. As we apply external field $h_z$, the asymptotic value of CR$_{y-y}$ is zero for FM $J$, indicating that the ground state is polarized in the $z$-direction. However, the asymptotic CR$_{y-y}$ for the large AFM $J$ is finite (surprisingly, -5/9 again) even for large external magnetic fields. This suggests that the ground state is still a quantum ground state, and this qualitatively marks the difference between the effect of FM and AFM $J$ to the $K$.
\\\\\indent
 For FM $J$, CR$_{y-y}$ first increases (opposite to the naive expectation) with $h_z$ being non-zero, which is remarkable. However it decreases as one increases the value of $h_z$. This suggests that for upto a critical field, the FM Heisenberg interaction is able to increase the CR$_{y-y}$ correlation as expected. For the intermediate strength of $J$, interesting variations of CR$_{y-y}$ appear.  As we increase the strength of the FM $J$, CR$_{y-y}$ starts to decrease and it happens at faster rate when magnetic field is stronger. This results in a critical $J_{c,fm}$ for each $h_z$ where CR$_{y-y}$ is maximum. We observe the plateau-like behaviour for the AFM $J$ for small values of $J$. The width of the plateaus increases as we increase the strength of the external magnetic field. It is important to note that this plateau-like behaviour persists even for the twelve-site cluster. Thus, our exact studies of correlation function uncover interesting aspects in comparison to earlier studies \cite{tikhonov-2011,song-2016,lunkin-2019,yang-2020,berke-prb-2020}.
\section{Expectation of plaquette conserved operator}\label{bpoperator}
\begin{figure}\centering
	\includegraphics[width=0.9\columnwidth,height=!]{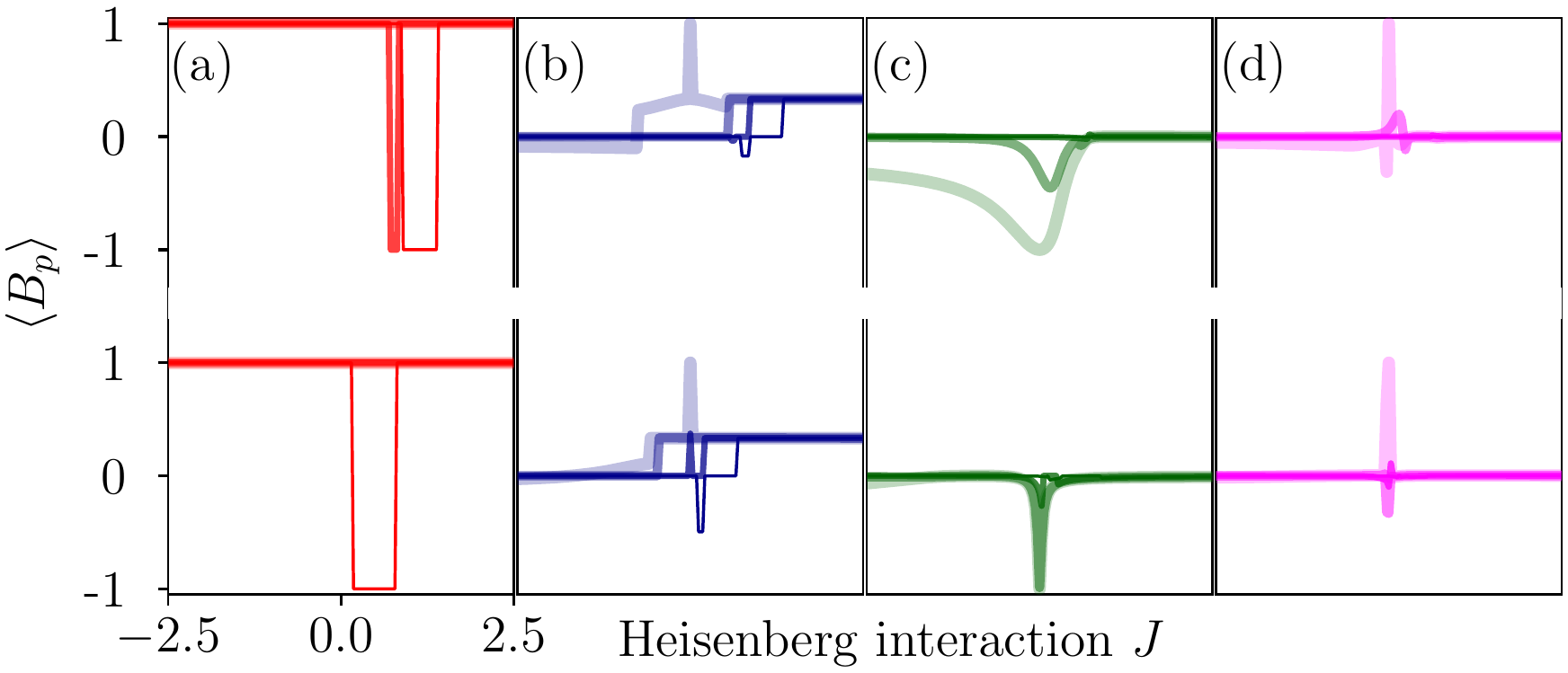}
	\caption{ Various plaquette operators are plotted against $J$, for different $h_z$. Red, blue, green, and magenta lines are for 4, 6, 8, and 12-site clusters. The lower panels correspond to AFM $K$, and the upper panels are for FM $K$. Lighter to darker shades indicate higher $h_z$ values; four different $h_z$ values have been used to plot: 0.0, 0.2, 1.0, and 2.0.}
	\label{bpfigure}
\end{figure}

In panel (a) of FIG.\ref{bpfigure}, we plot ground state expectation of four-sided $B_p$, $B_{1-4}= \langle \sigma^z_1 \sigma^z_2 \sigma^z_3 \sigma^z_4 \rangle$. Without $\Gamma$ interaction, $B_{1-4}$ is a conserved quantity as we have applied the magnetic field in $z$-direction. Thus, its expectation value needs to be $\pm 1$. We observe that at $J=0$, its value is $+1$ and remains so for FM $J$. However, for AFM $J$, its value changes to -1. After a critical $J_{afm}$, $B_{1-4}$ jumps back to 1. We observed that the range of $J$ for which it remains -1 coincides with the plateau region of the correlation function. This is because, in the previous section, we have seen that the ground state belongs to the odd-spin sector. We found that as the magnetic field is applied, the ground state wave function changes its parity, resulting in a change in net single spin flip in the $z$-direction and the sign change in $B_{1-4}$.
\\\\\indent
For six-site cluster, $B_p$ shows $+1$ for $J=0,~h_z=0$.
There are six four-site plaquettes and one six-site plaquette, and here we have plotted the $B_p$, which is associated with a four-sided plaquette ($B_p=\sigma_1^z\sigma_2^z\sigma_3^x\sigma_6^x$). Each two four-sided plaquette yields the six-sided plaquette. Due to the absence of reflection symmetry for the elementary square, we find the expectation value of the plaquette operator as +1 instead of -1 (which should be according to Lieb theorem~\cite{lieb-1994} if reflection symmetry is present). For nonzero $J$ and $K=1$, we observe a window of $J$ for which $B_p$ expectation takes a constant positive value and goes to zero after a critical FM $J$, whose magnitude increases with the external magnetic field. For FM $K$, this critical $J$ shifts to AFM $J$, suggesting an interesting competition between the relative sign of $J$ and $K$.
\\\\\indent
Having described four-site and six-site clusters in detail, we now explain the ground state $B_p$ expectation value for eight-site ($B_p=\sigma_1^z\sigma_2^y\sigma_7^x\sigma_6^z\sigma_5^y\sigma_4^x$) and twelve-site ($B_p=\sigma_1^z\sigma_2^y\sigma_9^x\sigma_8^z\sigma_7^y\sigma_6^x$) clusters. We note that elementary plaquettes in the eight-site cluster are four; hence, the Lieb theorem gives $B_p=-1$ for the ground state. We find that, for $J=0=h_z$, $B_p$ associated with a hexagonal plaquette is $-1$.
For $K=1$, any large values of $\pm J$, $B_p$ vanishes as expected, caused by mixing $B_p$'s all eigenstates in the presence of $J$. However, for small values of $\pm J$ and $K=1$, $B_p$ is large for small values of $h_z$. 
We note that some plaquette operators in the eight-site cluster contain only $\sigma^z$; hence, they commute with the external magnetic field. Thus, there are preferences for some eigenstates to be selected, and this causes finite large $B_p$ for small $\pm J$. The differences in FM $K$ compared to AFM $K$ are remarkable. For $K=-1$ with very large FM $J$, $B_p$ saturates at $\sim-0.4$ for $h_z=0$. The half-width maxima of $B_p$ for other $h_z$ is broader than that for AFM $K$. For the 12-site cluster, we observe that for $J=0, h_z=0$ and $K=\pm 1$ $B_p=1$ satisfying Lieb's theorem. For finite $h_z$ and AFM $K$, $B_p$ shows rapid oscillations (with positive and negative values of $B_p$) for a small $|J|$ window and vanishes afterwards. The same is true for FM $K$ though for small value of $h_z$, $B_p$ shows a saturation for large FM $J$ \cite{zhu-2018,han-li-2023,nandini-2019,liang-2018,yilmaz-2022}.
\section{Magnetization}\label{magnetization}
We now focus on some finite temperature properties of the clusters. Till now, we have found that various competing interactions $K$, $J$, $\Gamma$, and $h_z$ affect the zero temperature properties. Thus, it is natural to investigate some finite temperature properties to see the distinct ways each interaction affects them.
\begin{figure*}\centering
	\includegraphics[width=1.8\columnwidth,height=!]{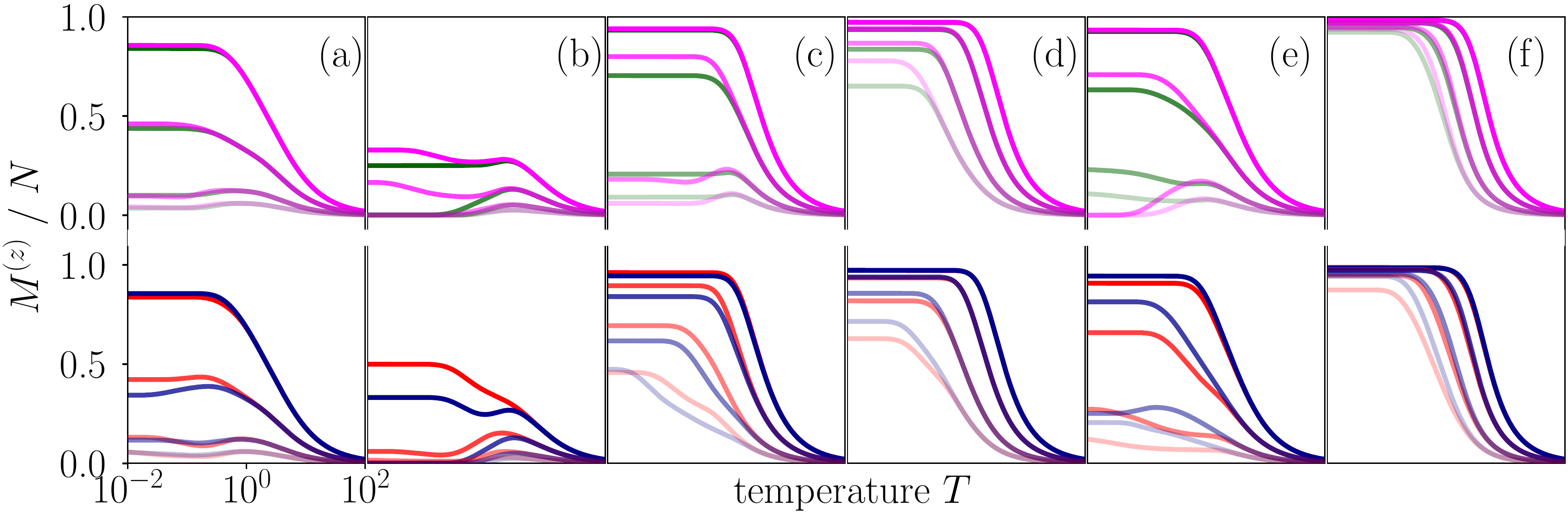}
	\caption{$M^{(z)}/N$ vs. temperature $T$, for different $J$ and $h_z$. $K$ is AFM in (a-c) and FM in (d-f). The value of $J$ is (a) 0.0, (b) 0.5, (c) -0.5, (d) 0.0, (e) 0.5, (f) -0.5. Red, blue, green, and magenta lines are for 4, 6, 8, and 12-site clusters. Lighter to darker shades indicate higher $h_z$ values; four different $h_z$ values have been used to plot: 0.2, 0.4, 1.0, and 2.0 .}
	\label{magnetizationfinitehz}
\end{figure*}
Magnetization in a finite field $h_z$ has been shown in FIG.\ref{magnetizationfinitehz}. It is normalized by the cluster size. Below, we describe some essential features.
\\\\\indent
{\it \small{FM ($K$=-1) vs. AFM Kitaev ($K$=1) for $J=0$}}: In the absence of $J$, we find there is a characteristic difference between the FM and AFM $K$ as evident from panel (a) and (d) of FIG.\ref{magnetizationfinitehz}. The dome-like structure
in the intermediate temperature appears only for AFM $K$. For FM $K$, magnetization's dome-like behaviour is absent, even for a small external magnetic field. The other significant difference is that the saturation magnetization for a given external magnetic field is higher for FM $K$ than the AFM one. Saturation magnetization is almost independent of the external magnetic field after it crosses a threshold value for FM $K$ in all clusters. However, for AFM $K$, the saturation magnetization greatly depends on the cluster size and gradually increases with cluster size.
\\\\\indent
{\it \small{FM ($K=-1$) vs. AFM Kitaev ($K=1$) for $J=0.5$}}: In the presence of AFM $J$, when $K=1$ (FIG.\ref{magnetizationfinitehz}, panel (b)), the dome-like structure in the magnetization is more pronounced and exists even for very high magnetic field. The most remarkable thing for $K=1$ is that, there are two limits of saturation magnetization $\mathcal{M}_1(h_1)$ and $\mathcal{M}_2 (h_2)$ where $\mathcal{M}_2 > \mathcal{M}_1 \ge 0$ with $h_2 > h_1$. We observed that the magnetic fields that yield a particular saturation value of magnetization at zero temperature belong to the same plateau region. We note that the qualitatively similar behaviour of magnetization has been found before experimentally~\cite{RColdea}.
For FM $K$ (FIG.\ref{magnetizationfinitehz}, panel (d)), the saturation magnetization is different for each magnetic field depending on the sign of $K$. It suggests that for $K$ positive, the nature of the ground state qualitatively changes for specific magnetic fields (which gives rise to a plateau in the correlation function as found in FIG. \ref{zzcorrelation}). However, for negative $K$, the ground states for different magnetic fields are smoothly related, reflecting the monotonous increase of saturation magnetization as we increase the magnetic field. 
\\\\\indent
{\it \small{FM ($K=-1$) vs. AFM Kitaev ($K=1$) for $J=-0.5$}}: For FM $J$, we observe the zero temperature magnetization limit for $K=1$ (FIG.\ref{magnetizationfinitehz}, panel (c)) depends on the external magnetic field and almost proportional to $h_z$. The dome-like structure exists, though it becomes weak. For FM $K$ (FIG.\ref{magnetizationfinitehz}, panel (f)), the zero temperature saturation magnetization is very high and independent of the external magnetic field after a critical $h_z$. The dome-like structure for the intermediate temperature range is absent. Thus, we see the effect of competing interactions manifested distinctly in the magnetization profile.
\\\\\indent
{\it \small{FM ($K=-1$) vs. AFM Kitaev ($K=1$) for $\Gamma=\pm0.5$}}: 
For this, we refer to FIG.\ref{magnetization_finite_Gamma} in Appendix \ref{App:A}. The zero field magnetization is very small ($\sim 10^{-8}$) compared to the finite field value. Secondly, for AFM $K$ (panel (a) and (b)), the saturation magnetization at zero temperature increases almost proportionally with the applied magnetic field, whereas, with FM $K$ (panel (c) and (d)), magnetization saturates to close to 1 for higher values of different magnetic fields. Interestingly, the dome-like structure is completely absent for FM $K$ (panel (c) and (d)) but appears for AFM $K$ (and for very weak magnetic field $h_z\sim 0.2$). The saturation magnetization becomes almost independent of cluster size for large magnetic fields, but for smaller magnetic fields, the saturation magnetization is entirely cluster-dependent.
\section{Susceptibility}\label{susceptibility}
\begin{figure*}
	\includegraphics[width=1.8\columnwidth,height=!]{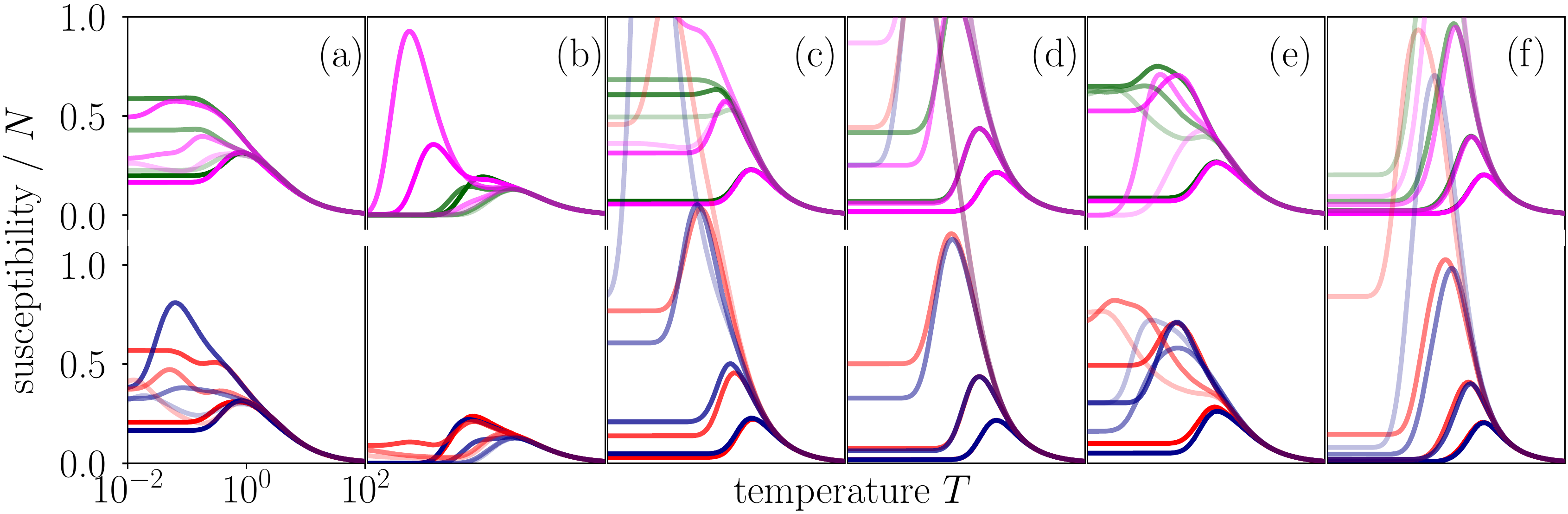}
	\caption{Susceptibility / $N$ vs. temperature $T$, for different $J$ and $h_z$. $K$ is AFM in (a-c) and FM in (d-f). The value of $J$ is (a) 0.0, (b) 0.5, (c) -0.5, (d) 0.0, (e) 0.5, (f) -0.5. Red, blue, green, and magenta lines are for 4, 6, 8, and 12-site clusters. Lighter to darker shades indicate higher $h_z$ values; four different $h_z$ values have been used to plot: 0.2, 0.4, 1.0, and 2.0 .}
	\label{susceptibilitygamma0}
\end{figure*}

We now proceed to discuss the susceptibility. As before, we find characteristic signatures of competition among different interactions. In FIG.\ref{susceptibilitygamma0}, we have plotted susceptibilities for various clusters and different combinations of parameters. We have investigated susceptibility for finite magnetic field $h_z$ from $0.2$ to $2.0$ (in step of $0.2$). However, we have only plotted for $h_z=0.2,0.4,1.0,2.0$ to describe how susceptibilities vary with magnetic fields. We first begin with a comparison of susceptibility for $J=0$.
\\\\\indent
{\it \small{FM ($K=-1$) vs. AFM Kitaev ($K=1$) for $J=0$}}:
First, we describe how susceptibility varies when $J$ is absent and the differences between FM and AFM Kitaev interactions. As we see from panel (d) of FIG.\ref{susceptibilitygamma0}, susceptibilities for different clusters with FM $K$ are similar. The susceptibility is zero at very high temperatures and maximum at some intermediate values and becomes constant as the temperature tends to zero. As expected, the peak height gradually decreases with increasing magnetic field and appears at higher temperatures. The saturation susceptibility also depends on the magnetic field and decreases with increasing value of $h_z$. However, for AFM $K$ (FIG.\ref{susceptibilitygamma0}, panel (a)), an external magnetic field's effect on susceptibilities is completely different. First, there is no divergence in the susceptibilities for any (non-zero) magnetic field observed for FM $K$. Also, susceptibility behaviour is not uniform for different magnetic fields over the cluster sizes. For example, the susceptibility can be high or low in the twelve-site cluster at a given temperature, depending on the external magnetic field. Identical trends can be seen for the eight-site cluster. Though there are peaks in the susceptibilities at a given magnetic field for the twelve-site cluster, for the eight-site cluster, peaks are very mild and only appear for a small magnetic field. Owing to the small sizes and intricate nature of the ground state, the susceptibility for four and six-site clusters show complex behaviour.
\\\\\indent
{\it \small{FM ($K=-1$) vs. AFM Kitaev ($K=1$) for $J=0.5$}}: An AFM $J$ brings in a significant difference to the susceptibilities of FM $K$ (FIG.\ref{susceptibilitygamma0}, panel (e)); the divergences found for $J=0$ at a low magnetic field now appear at some finite magnetic field for the twelve-site cluster. On the other hand, there are no pronounced maxima at some magnetic fields for the four and eight-site clusters, but such pronounced maxima appear for the six-site cluster. Similarly, we find, though, in the absence of $J$, there is no pronounced maxima for FM $K$, an AFM $J$ causes susceptibility to have pronounced maxima (FIG.\ref{susceptibilitygamma0}, panel (b)). Also, at low temperatures, the susceptibilities converge to very low values for AFM $K$, but for FM $K$, it saturates to quite large values. Thus, we see clear evidence of competing effects on the susceptibilities due to the $J$ and $K$.
\\\\\indent
{\it \small{FM ($K=-1$) vs. AFM Kitaev ($K=1$) for $J=-0.5$}}: In the presence of FM $J$, we observe that, for $K=-1$ (FIG.\ref{susceptibilitygamma0}, panel (f)), the peak structure re-appears in susceptibilities and it is divergent mostly for a small magnetic field. As we increase the magnetic field, the peak height gradually decreases across all clusters. They tend to saturate at zero for intermediate to large magnetic fields at very low temperatures. For the AFM $K$ (FIG.\ref{susceptibilitygamma0}, panel (c)), the pronounced peak appears for the twelve-site cluster at some intermediate critical magnetic field. Except for this critical magnetic field, the susceptibilities are generally regular. However, the susceptibility value is not monotonously increasing or decreasing at a given temperature on the external magnetic field. For four and six-site clusters, pronounced peaks re-appear at very small magnetic fields, but for the eight-site cluster, there is no peak-like structure; instead, there is a mild dome-like structure.
\\\\\indent
{\it \small{FM ($K=-1$) vs. AFM Kitaev ($K=1$) for $\Gamma=\pm0.5$ at $J=0$}}: Now we discuss the effect of $\Gamma$ on susceptibility in FIG.\ref{susceptibility_finite_Gamma} of \ref{App:A}. We observe that there are some cases where susceptibility diverges at zero temperature, such as for the four site cluster, for $K=\pm1,~\Gamma\pm0.5$, at zero magnetic field and also for eight site cluster for $K=-1,~\Gamma=-0.5,~h_z=0$. Otherwise, susceptibility is finite at zero temperature. There is a clear distinction in behaviour between FM and AFM Kitaev interactions. Interestingly, the susceptibility for the six-site cluster looks completely different than other clusters. In short, we find that the effect of finite $\Gamma$ on susceptibility is non-monotonous for the AFM Kitaev model as a function of temperature and magnetic field.
\section{Specific heat}\label{specificheat}
We now proceed to analyze specific heat for the $K$-$J$-$\Gamma$ systems. Much of the recent experimental results invested a great deal of analysis of the double-peak structure of the Kitaev system. The lower temperature peak is associated with a long-range ordered state, and the high-temperature peak is attributed to a deconfined spin-liquid phase~\cite{Loidl-2021}. As we will observe, depending on the relative strength of $K$, $J$, $\Gamma$, and $h_z$, there exists a well-established two-peak structure, a single sharp peak or a wide broad peak. On the other hand, for a pure Kitaev model, the peak at lower temperatures is due to the excitations in the flux sectors. Below, we systematically present the specific heat for various parameter combinations.
\begin{figure*}
	\includegraphics[width=1.8\columnwidth,height=!]{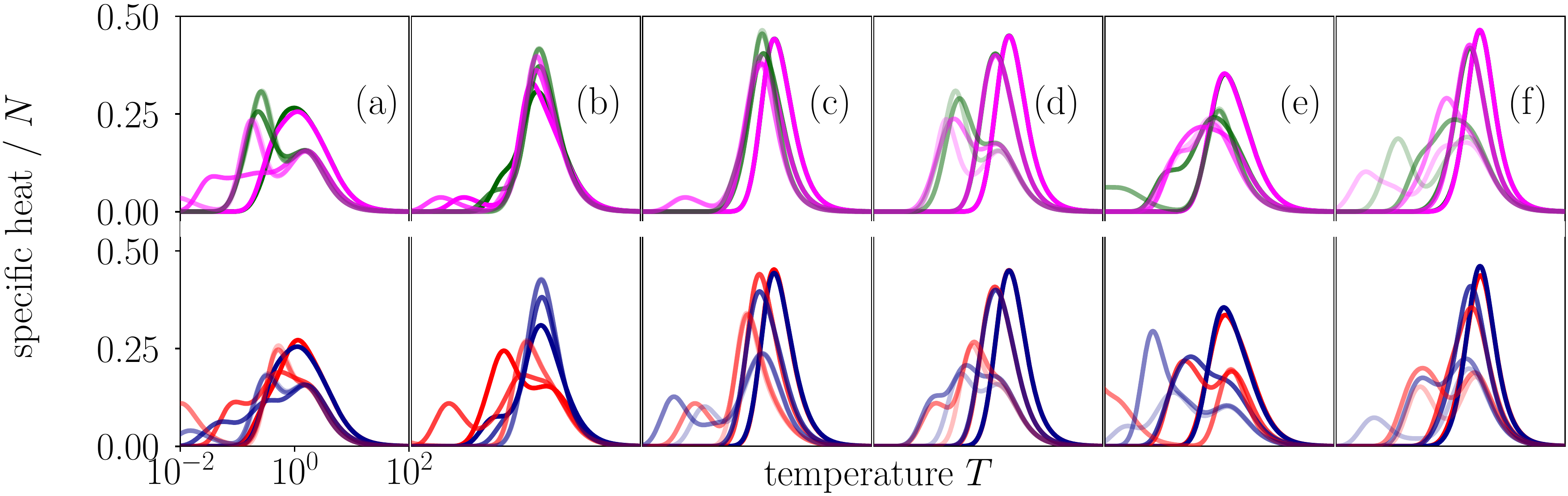}
	\caption{Specific heat /$N$ vs. temperature $T$ for different $J$ and $h_z$ values, in absence of $\Gamma$. $K$ is AFM in (a-c) and FM in (d-f). The value of $J$ is (a) 0.0, (b) 0.5, (c) -0.5, (d) 0.0, (e) 0.5, (f) -0.5. Red, blue, green, and magenta lines are for 4, 6, 8, and 12-site clusters. Lighter to darker shades indicate higher $h_z$ values; four different $h_z$ values have been used to plot: 0.0, 0.2, 1.0, and 2.0 .}
	\label{specificheatgammazero}
\end{figure*}
\\\\\indent
{\it \small{FM ($K=-1$) vs. AFM Kitaev ($K=1$) for $J=0$}}: The primary difference between FM and AFM $K$ in the absence of $J$ (FIG.\ref{specificheatgammazero}) is that the two peaks in the case of AFM $K$ (panel (a)) are more frequent for many values of the external magnetic field, whereas for FM $K$ (panel (d)), the two peak structure only appears for small magnetic field. This implies that the exact quantum eigenstates are renormalized for $K=1$ and $K=-1$ differently under the external magnetic field. The frequent occurrences of two peak structures imply that the eigenstates are mildly modified in the presence of an external magnetic field in the case of AFM $K$. On the other hand, for FM $K$, the effect of an external magnetic field is stronger in destroying the two-peak structure, implying confined phases. The broader wide peaks in the case of AFM $K$ strengthen this understanding further that the two-peak structure mostly survives, denoting the robustness of the deconfined phase in an external magnetic field. With this background understanding, we now explain what happens when we consider a finite AFM $J$.
\\\\\indent
{\it \small{FM ($K=-1$) vs. AFM Kitaev ($K=1$) for $J=0.5$}}: In the case of AFM $K$, the effect of having $J=0.5$ (FIG.\ref{specificheatgammazero}, panel (b)) is that it makes the two peak structure remain intact but with a qualitative difference. Also, the distance between the peaks has increased for $J=0.5$. This denotes that AFM $J$ stabilizes the eigenspectrum determined by the conserved flux sector qualitatively in a magnetic field. Remarkably, for FM $K$ (FIG.\ref{specificheatgammazero}, panel (e)), we see the re-entrance of a two-peak structure in the presence of AFM $J$. We thus conclude that an AFM $J$ helps stabilize the KSL phase compared to the absence of $J$ in the presence of an external magnetic field.
\\\\\indent
{\it \small{FM ($K=-1$) vs. AFM Kitaev ($K=1$) for $J=-0.5$}}: For FM $J$, we observe the two-peak structure rarely occurs for AFM $K$ (FIG.\ref{specificheatgammazero}, panel (c)) in comparison to $J=0.5$. However, the exact pattern of specific heat also depends on the cluster sizes. Most of the clusters have a single peak for high $h_z$. Similarly, we observe mostly single peaks in the case of FM $K$ (panel (f)), with larger $h_z$. This tells us that an FM $J$ works against the KSL phase. It is interesting to note that an AFM $J$, in general, induces a tendency to have a quantum paramagnet state which seems to favour a KSL phase in comparison to an FM $J$, which induces a collinear order in the system, which seems to destabilize the KSL state.
\\\\\indent
{\it \small{FM ($K=-1$) vs. AFM Kitaev ($K=1$) for $\Gamma=\pm0.5$ at $J=0$}}: Now we discuss the effect of $\Gamma$ interaction on specific heat; for simplicity, we only describe in detail the case for $J=0$. We refer to FIG.\ref{specific_heat_finite_Gamma} in Appendix \ref{App:A}. There is a clear difference between the AFM (panel (a), (b)) and the FM (panel (c), (d)) Kitaev case. In the former case, the double peak structure is either sustained or gets broadened by applying a magnetic field, while in the latter case, when we introduce $h_z$, the double peak structure merges into a single peak. This signifies that AFM $K$ is more robust against $\Gamma$ interactions (in the presence of a magnetic field) to keep its spectrum closer to the one obtained for $\Gamma=0$. Surprisingly, high field characteristics of specific heat at a particular parameter value are almost cluster-independent. This shows that in the presence of a strong magnetic field, the distribution of energy eigen spectra is very similar for all clusters. Interestingly, only in $K=1,~\Gamma=-0.5$, a triple peak structure arises for some magnetic field values in all clusters.\\\\
\indent
The two peak structure of specific heat has been acknowledged as confirmation of the Kitaev spin liquid phase both theoretically and experimentally \cite{kataoka-jpsj-2020,andrade-2020,tathagata-2020,bachus-2021,balz-prb-2021,feng-2020,yogesh-2017,nasu-2015,yamaji-2016}. The effect of magnetic field on magnetization, susceptibility, and specific heat resonate with some recent studies. For example, the magnetic field is found to stabilize the magnetically ordered state before melting into a quantum paramagnetic state in $\alpha\mhyphen{\rm RuCl}_3$~\cite{balz-prb-2021} for FM $K$ and $J$. In accordance with this, we find from panel (f) of FIG.\ref{specificheatgammazero} ($K$ and $J$ are both FM), only the double peak structure is present for zero or very small magnetic fields. An increased magnetic field stabilizes the one-peak structure across all clusters. Similarly, for susceptibility, in accordance with the previous studies~\cite{andrade-2020}, we find, in some cases, the susceptibility is maximum at some intermediate magnetic field and not at zero fields (see FIG.\ref{susceptibilitygamma0}). Although the combinations of $\left(J, K\right)$ are different from those taken in ~\cite{andrade-2020} where $J$ and $K$ both are taken to be FM and contain other complex interactions, it is important that the small clusters can capture the qualitative essence\cite{kataoka-jpsj-2020}. Similarly, in another previous study~\cite{yamaji-2016}, which considered a strong FM $K$ and a weak AFM $J$ in the presence of next-nearest neighbour interaction, the specific heat is shown to change the width and height of the two peaks depending on the external magnetic field. These scenarios can be compared with FIG.\ref{specificheatgammazero}, where across all the clusters, similar behaviour is observed. We comment that due to the limited cluster size, the effect of next-nearest neighbour interaction is simulated easily.
\section{Discussion}\label{discussion}
We now summarize our extensive investigation of the Kitaev-Hesineberg-$\Gamma$ model in finite-size clusters. Starting with the pure-Kitaev model, when $J$ is added on top of it after a critical value of $J$, the eigenspectrum of the system varies linearly and merges to the classical states for FM $J$ or appropriate singlet states for AFM $J$. In the large FM limit, the energy density is given by $E=-6J+2K$ independent of the cluster size. However, very interestingly, in the large AFM $J$, the asymptotic energy density depends on the cluster size, implying the non-trivial presence of Kitaev and $\Gamma$ interactions. Near the Kitaev limit, some branches of eigenvalues bend and do not change signs across zero as $J$ changes from negative to positive. Also, there are many eigenenergies whose values vary linearly and change signs when the sign of $J$ is reversed. In the presence of a magnetic field, for FM $K$, the effect is similar for all clusters. However, for AFM $K$, the effect is shown to depend largely on the cluster size. This follows from the fact that AFM $J$ induces significant quantum fluctuations, which depend greatly on cluster size.
\\\\\indent
Next, we investigated the nearest-neighbour correlation function on a `$z$-type' bond. The correlation function exhibits a plateau for AFM $J$ in the presence of the external magnetic field. This happens as the exact ground state becomes a simultaneous eigenstate of the Heisenberg and Zeeman's terms. On the other hand, for FM $J$, the correlation function shows non-monotonous behaviour as a consequence of competition between Kitaev and Heisenberg interaction. This non-monotonicity is found to be enhanced for larger clusters and hence could be of relevance to real material. The fact that for AFM $J$ and external magnetic field, the ground state becomes a simultaneous eigenstate of the Heisenberg and Zeeman field could lead to greater insight into forming a variational wave function for the Kitaev-Heisenberg-$\Gamma$ system in the thermodynamic limit.
\\\\\indent
Next, we analyzed the effect of FM and AFM $K$ and external magnetic field on magnetization, susceptibility, and specific heat for various $J$. Magnetization shows a dome-like pattern for low-field and AFM $K$, absent for FM $K$. Also, zero-temperature magnetization has a large dependency on the external magnetic field. It can be evenly spaced from zero to a maximum or close to unity (when normalized by the system size), depending on whether $K$ is AFM or FM. $J$ modulates the magnetization's qualitative nature, showing distinct quantum effects depending on its sign. Since we found plateau regions in correlation data where the ground state does not change for a range of magnetic fields, magnetization at zero temperature saturates to a particular value for all the magnetic field values in that range. A similar intriguing effect of various exchange couplings and external magnetic fields on susceptibility and specific heat is observed.
\\\\\indent
With the recent intense effort to understand the $K$-$J$-$\Gamma$ system, we think our extensive analysis will help quantify the role of quantum fluctuations in these systems in the thermodynamic limit. Further, a study of the classical ground state on the Kitaev cluster exhibits interesting order by singularity mechanism\cite{sarvesh-2020,anjana-2018}. Extending this study to larger clusters and in the presence of $J$ could provide valuable insights to explore new mechanisms of quantum fluctuation~\cite{khatua-2021}. Further within the slave-fermion language, implementing a PSG calculation\cite{wen-2002,yasir-2020,atanu-2023} for this finite size system utilizing various symmetries available and their response to various non-Kitaev interactions could be useful to compare the exact analysis done here, and it will be taken as a future study.
\section*{Acknowledgement} 
S.M.P. acknowledges SAMKHYA (High-Performance Computing facility provided by the Institute of Physics, Bhubaneswar) for numerical computation. S.M.P. also thanks Arnob Kumar Ghosh for many valuable discussions. S.M. acknowledges support from ICTP through the Associate's Programme (2020-2025). S.M. also thanks G. Baskaran and Nicola Seriani for interesting discussions.
\bibliography{bibfile}{}
\section{Appendix}
\label{appendix}
\appendix
\newcounter{defcounter}
\setcounter{defcounter}{0}
\vspace{20pt}
\section{Finite $\Gamma$ results}{\label{App:A}}
\begin{figure}[H]
	\includegraphics[width=0.9\columnwidth,height=!]{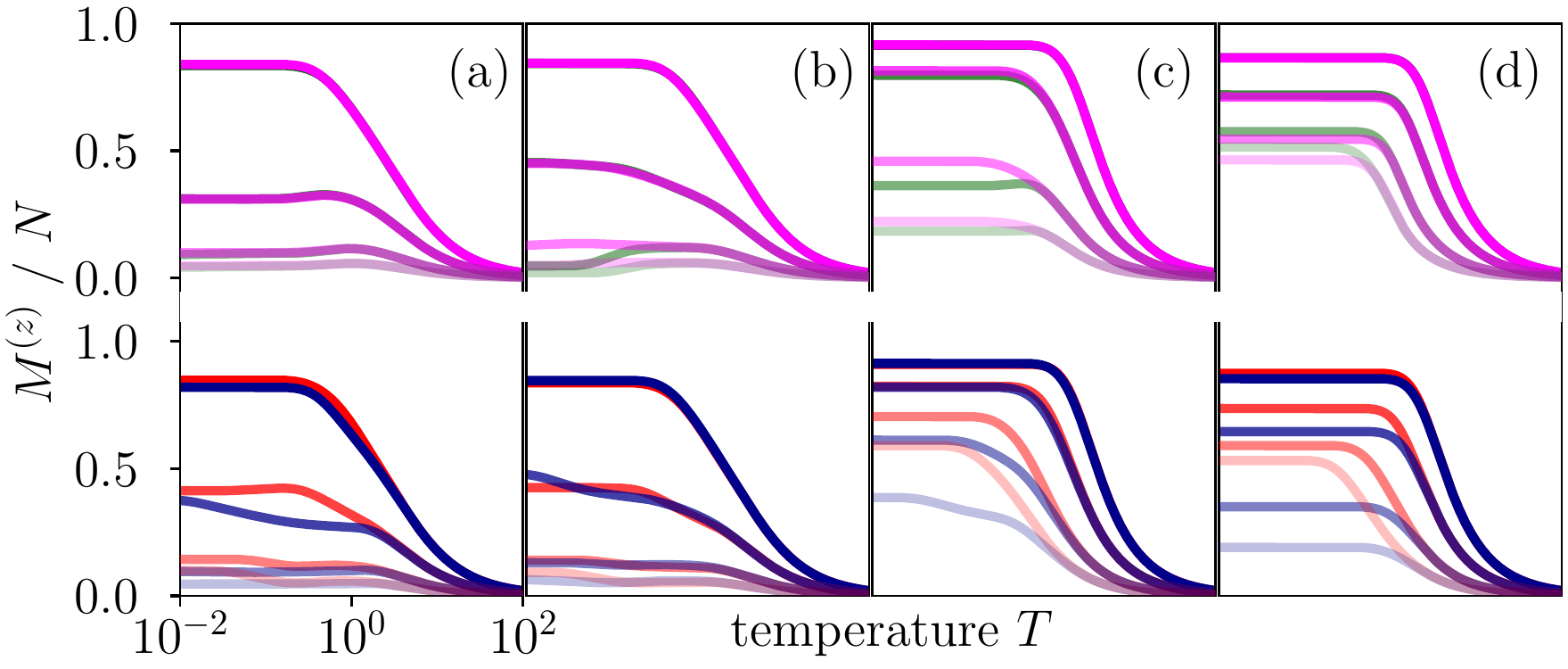}
	\caption{$M^{(z)}/N$ vs. temperature $T$, in absence of $J$. $K$ is AFM in (a-b) and FM in (c-d). Value of $\Gamma$ is (a) 0.5, (b) -0.5, (c) 0.5, (d) -0.5 . Line coloured in red, blue, green, and magenta are for 4, 6, 8, and 12-site clusters. Lighter to darker shades indicate higher $h_z$ values; four different $h_z$ values have been used to plot: 0.2, 0.4, 1.0, and 2.0 .}
	\label{magnetization_finite_Gamma}
\end{figure}
\begin{figure}[H]
	\includegraphics[width=0.9\columnwidth,height=!]{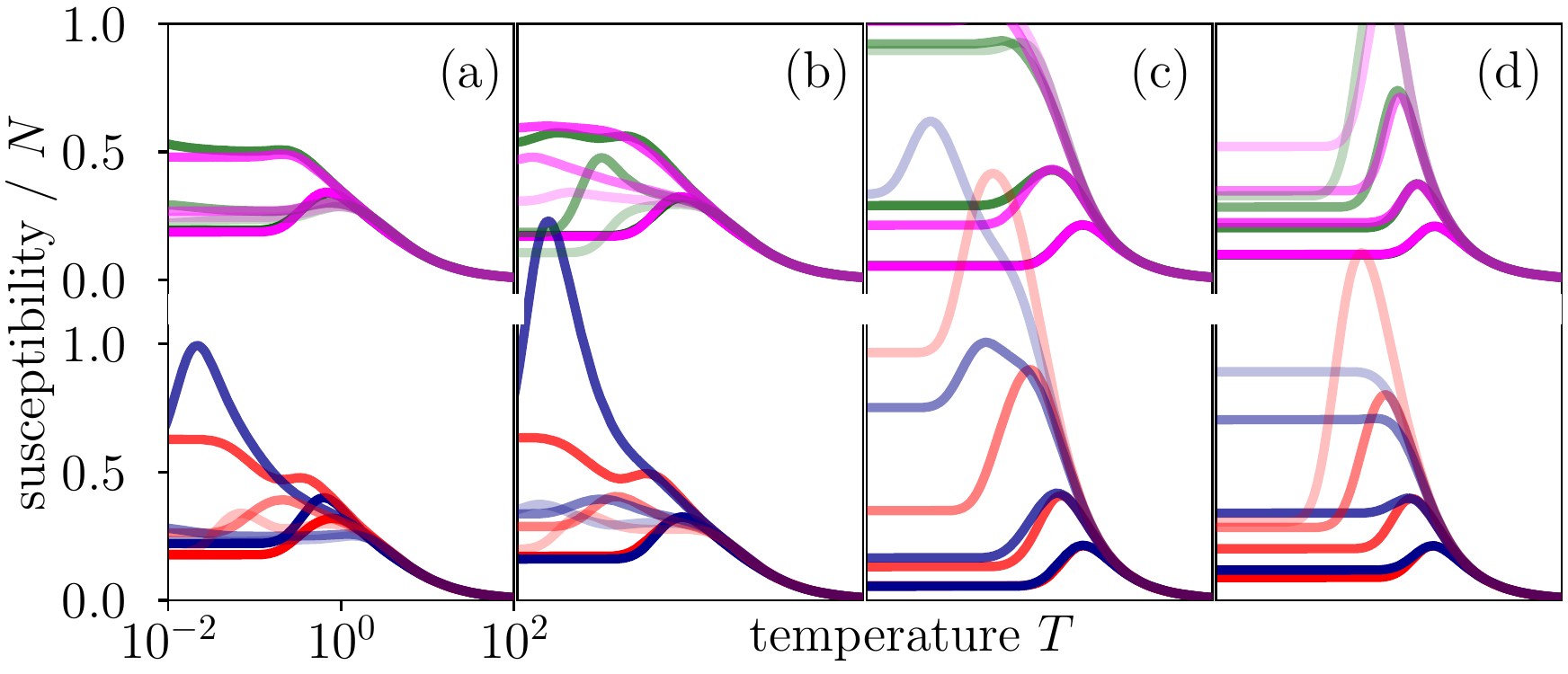}
	\caption{Magnetic susceptibility / $N$ vs. temperature $T$, in absence of $J$. $K$ is AFM in (a-b) and FM in (c-d). Value of $\Gamma$ is (a) 0.5, (b) -0.5, (c) 0.5, (d) -0.5 . Line coloured in red, blue, green, and magenta are for 4, 6, 8, and 12-site clusters. Lighter to darker shades indicate higher $h_z$ values; four different $h_z$ values have been used to plot: 0.2, 0.4, 1.0, and 2.0 .}
	\label{susceptibility_finite_Gamma}
\end{figure}
\begin{figure}[H]
	\includegraphics[width=0.9\columnwidth,height=!]{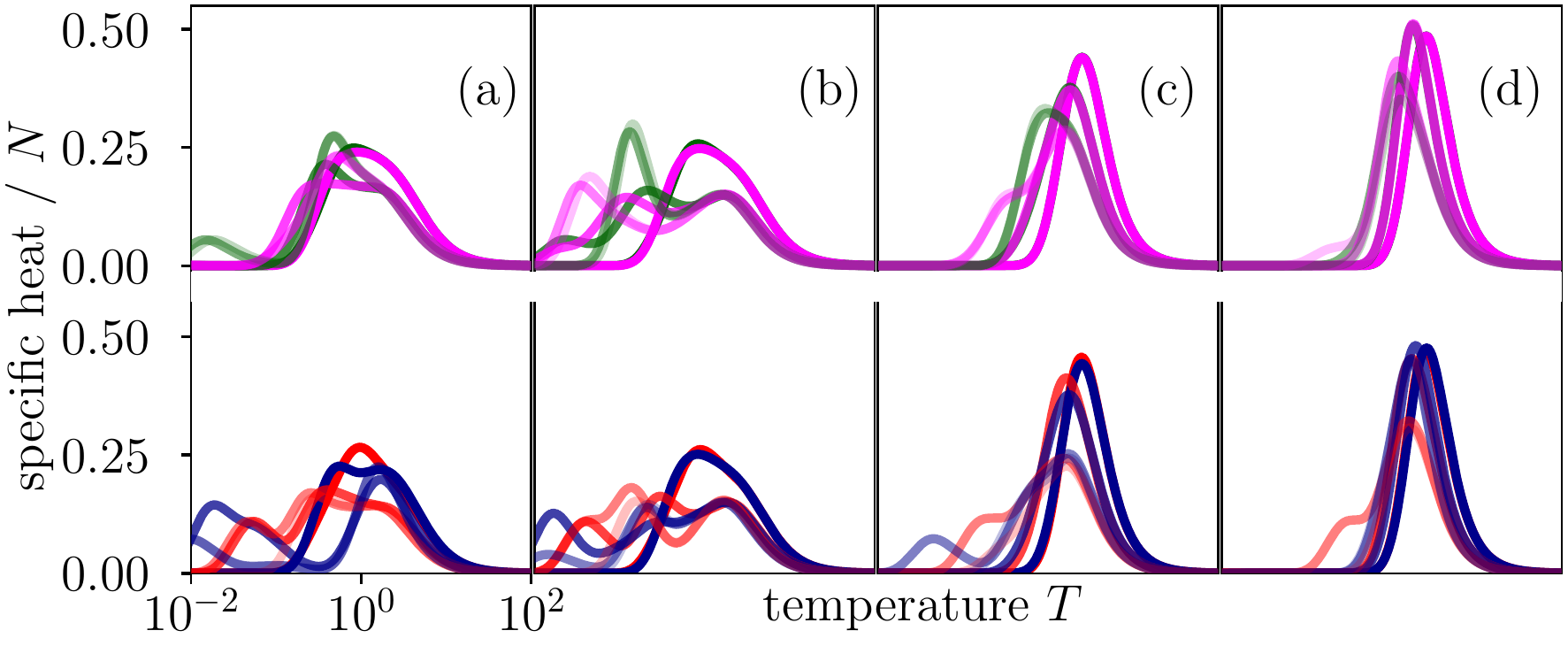}
	\caption{Specific heat /$N$ vs. temperature $T$, in absence of $J$. $K$ is AFM in (a-b) and FM in (c-d). Value of $\Gamma$ is (a) 0.5, (b) -0.5, (c) 0.5, (d) -0.5 . Line coloured in red, blue, green, and magenta are for 4, 6, 8, and 12-site clusters. Lighter to darker shades indicate higher $h_z$ values; four different $h_z$ values have been used to plot: 0.0, 0.2, 1.0, and 2.0 .}
	\label{specific_heat_finite_Gamma}
\end{figure}
\end{document}